\documentclass[usenatbib]{mnras}

\usepackage{newtxtext,newtxmath}

\usepackage[T1]{fontenc}
\usepackage{ae,aecompl}
\usepackage{tabularx}
\usepackage{threeparttable}

\usepackage{graphicx}	
\usepackage{amsmath}	
\usepackage{amssymb}	

\newcommand{\Msun}{{\rm M}_{\odot}}

\newcommand{\kms}{\rm km \, s^{-1}}
\newcommand{\Vpeak}{V_{\rm peak}}

\DeclareMathOperator\erf{erf}
 
\title[]{How low does it go?  Too few Galactic satellites with standard reionization quenching}
\author[A. S. Graus et al.]{Andrew S. Graus$^{1,2}$,\thanks{E-mail: agraus@uci.edu}
James S. Bullock$^{1}$,
Tyler Kelley$^{1}$, 
Michael Boylan-Kolchin$^{2}$,  \newauthor
Shea Garrison-Kimmel$^{3}$, 
 Yuewen Qi$^{1}$
\\
$^{1}$Center for Cosmology, Department of Physics and Astronomy, 4129 Reines Hall, University of California Irvine, CA 92697, USA \\
$^{2}$The University of Texas at Austin, Department of Astronomy, 2515 Speedway, Stop C1400, Austin, TX 78712-1205\\
$^{3}$TAPIR, Mailcode 350-17, California Institute of Technology, Pasadena, CA 91125, USA \\
}

\date{Accepted XXX. Received YYY; in original form ZZZ}

\pubyear{2018}

\begin{document}
\label{firstpage}
\pagerange{\pageref{firstpage}--\pageref{lastpage}}
\maketitle

\begin{abstract}
A standard prediction of galaxy formation theory is that the ionizing background suppresses galaxy formation in haloes with peak circular velocities smaller than $\Vpeak \simeq 20 \,\kms$, rendering the majority of haloes below this scale completely dark. We use a suite of cosmological zoom simulations of Milky Way-like haloes that include central Milky Way disk galaxy potentials to investigate the relationship between subhaloes and ultrafaint galaxies.   We find that there are far too few subhaloes within 50 kpc of the Milky Way that had $\Vpeak \gtrsim 20\,\kms$  to account for the number of ultrafaint galaxies already known within that volume today.  In order to match the observed count, we must populate subhaloes down to $\Vpeak \simeq 6\,\kms$ with ultrafaint dwarfs.  The required haloes have peak virial temperatures as low as $1,500$ K, well below the atomic hydrogen cooling limit of $10^4$~K.   
Allowing for the possibility that the Large Magellanic Cloud contributes several of the satellites within 50 kpc could potentially raise this threshold to $10\,\kms$ ($4,000$ K), still below the atomic cooling limit and far below the nominal reionization threshold.
\end{abstract}




\section{Introduction} \label{s:intro}

One of the foundational developments in near-field cosmology was the discovery of ultrafaint dwarf galaxies in the Sloan Digital Sky Survey (SDSS) (see \citealt{Willman2010} for a review).
More recent efforts from DES, PanSTARRS, and MagLiteS (among other surveys) have led to many additional discoveries of ultrafaint Milky Way satellites \citep{Koposov15,ADW15,Laevens15a,Laevens15b,ADW16}; the current census of ultrafaint satellites in the Milky Way is approximately 45. These galaxies are incredibly faint, with luminosities as low as $\rm \sim 350\,L_\odot$, and heavily dark matter dominated \citep{Simon07}. As such, they may represent the long-discussed `Missing Satelltes' of the Milky Way \citep{Klypin1999,Moore99}. We expect many more such objects to exist within the virial radius of the Milky Way; only about half the sky has been surveyed and is only complete to within $\sim 30$ kpc for the faintest dwarfs \citep[][]{Walsh07}.
 
The stars in all ultrafaint galaxies are universally old ($\ga 11 \,{\rm Gyr}$) and this lends credence to the idea that their star formation was shut down in response to reionization at high redshift \citep{Bovill09,Brown14,Weisz14,Wheeler15}. While most of these ancient ultrafaint dwarfs are satellites of larger systems, it is statistically unlikely that environmental quenching could have quenched star formation early enough in these objects to explain the absence of young stars in all of them \citep{KRW18}.  
 
Reionization suppression is an attractive mechanism for explaining the uniformly ancient stellar populations of ultrafaint dwarfs, as there should be a dark matter halo mass scale below which galaxy formation is severely limited by the ionizing background \citep{Efstathiou92}.
 The majority of models that have explored the reionization suppression scale have found that most dark matter haloes with peak maximum circular velocities ($\Vpeak$) smaller than $\Vpeak \simeq 20-30 \,\kms$ are unable to accrete gas after reionzation \citep{Thoul96,Gnedin00,Hoeft06,Okamoto08}.  This is not unexpected, as haloes of this size have virial temperatures of $T_{\rm vir} \sim 20,000$ K, which is similar to the IGM temperature after reionization \citep[e.g.][]{McQuinn16,Onorbe17}. 
Suppression at this scale also naturally solves the missing (classical) satellites problem  \citep{Bullock00,Benson02,Somerville02,Kim17,Read18}.

More recently, \citet{Ocvirk16} have used full radiative transfer simulations of the Local Group to show that reionization suppresses galaxy formation in haloes with $M_{\rm vir}$ $\simeq$ $5\times 10^{8}$ $\rm M_{\odot}$ measured at $\it{z}$ = 5.5, which is equivalent to $V_{\rm max}$ $\simeq$ 20-25 $\rm km \,s^{-1}$ at this redshift~\footnote{Note that $V_{\rm vir} \propto (1+z)^{1/2}$ at fixed halo mass if we ignore the potential (mild) evolution in the virial overdensity definition.}. A similar quenching threshold is seen in high-resolution hydrodynamic simulations that track dwarf galaxy formation down to redshift zero \citep{Sawala16b,Munshi17,Benitez17,Fitts17,Maccio17}. For example, \citet{Fitts17} have used FIRE zoom simulations to study dwarf galaxy formation and find that the majority of haloes with peak subhalo masses below $10^9\,\Msun$ form no stars.  This is equivalent to a threshold at $V_{\rm peak} \simeq 20\,\kms$.  

A second scale of relevance for low-mass galaxy formation is the atomic hydrogen cooling limit at $10^4$ K, which corresponds to a $V_{\rm peak} \simeq 16\,\kms$ halo. Systems smaller than this would require molecular cooling to form stars.  Taken together, one might expect that most ultrafaint satellite galaxies of the Milky Way should reside within subhaloes that fell in with peak circular velocities in the range $16-30\,\kms$, though these systems would have lower maximum circular velocities ($V_{\rm max}$) {\em {today}} as a result of dark matter mass loss after infall onto the Milky Way's halo ($V_{\rm max} \le V_{\rm peak}$).

In addition to tidal stripping, the destruction of dark matter subhaloes due to interactions with the potential of the central galaxy itself is a crucial physical process that must be included in any comparison to satellite galaxy counts \citep{D10,Brooks14,Zhu2016,Wetzel2016,SGK2017}. The effect of the central galaxy decreases subhalo abundances by about a factor of two within the virial radius compared to dark matter only simulations; a similar effect is seen in massive elliptical galaxy haloes \citep{Despali17,Graus18}. The enhanced destruction is particularly important for subhaloes close to the central galaxy.  At the Milky Way scale, almost all haloes above the resolution limit are destroyed within 20 kpc \citep{SGK2017}.  This level of depletion, combined with the pace of new discoveries at small Galacto-centric radii, leads us to ask whether there may be \textit{too many} Milky Way satellites rather than not enough \citep[see, e.g.][]{Jethwa18,Li18}.

In this work, we use a new suite of cosmological zoom simulations of Milky Way-like haloes simulated with an evolving Milky Way disk (plus bulge) potential to explore the relationship between dark matter haloes and ultrafaint galaxies.  We show that a conventional reionization suppression scale at $\Vpeak \simeq 20\,\kms$ drastically under-produces the count of Milky Way satellites within $50$ kpc of the Galactic center.  Assuming the Milky Way is typical of our simulation suite, it appears that a significant fraction of ultrafaint galaxies must form in subhaloes with peak circular velocities less than $10\,\kms$. These haloes have virial temperatures of $T_{\rm vir}$ $< 4,000$ K, which is well below the nominal atomic hydrogen cooling limit. In section, \ref{s:sims and methods} we describe the simulations. Section \ref{s:results} presents our results. Section \ref{s:caveats} discusses some possibilities that could alter our conclusions and Section \ref{s:conclusion} provides a summary discussion.

\section{Simulations and Methods} \label{s:sims and methods}

In this work, we use a new zoom simulation suite performed by Kelley et al.~(in preparation) that consists of dark matter only Milky Way-like haloes simulated with a disk + bulge potential to account for the destructive effects of the central galaxy.  The galaxy potentials were implemented in a manner similar to that discussed in \citet{SGK2017}, who showed that this approach mimics the enhanced subhalo destruction seen in full hydrodynamics simulations. 

\begin{figure*}
	\includegraphics[width=0.49\textwidth, trim = 0 0 0 0]{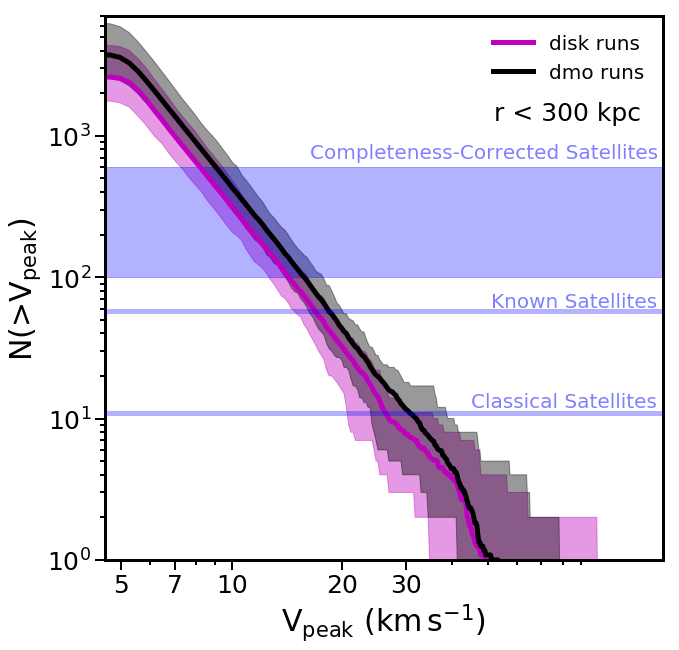} 
    	\includegraphics[width=0.49\textwidth, trim = 0 -10 0 0]{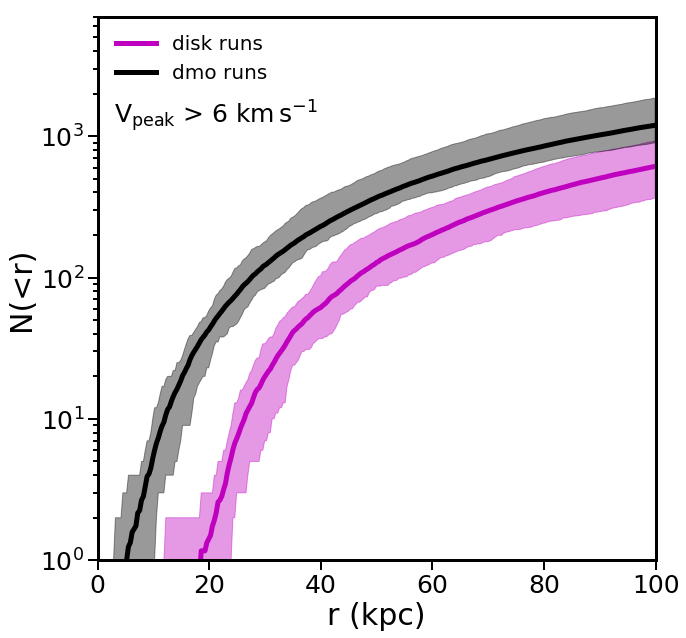}
	\centering
	\caption[Subhalo props]{
    \textit{Left:} Cumulative subhalo counts within 300 kpc as a function of the peak maximum circular velocity achieved over subhalo's history, $V_{\rm peak}$. The black distribution represents the full range of our dark matter only simulations, while the magenta distribution represents the simulations with analytic disk potentials.  The solid lines are medians.   The horizontal lines show the number of classical satellites, All presently known satellites, and the range of total satellites expected based on sky and volume completeness corrections.  \textit{Right:} Radial distribution of all subhaloes with $V_{\rm  peak}$ $\geq 6 \,\kms$ for the dmo simulations (black) and the dmo+disk simulations (magenta). }
	\label{fig:subhalo_props}
\end{figure*}

The simulations were run with GIZMO \citep{Hopkins15}, which uses an updated version of the TREE+PM gravity solver from \texttt{GADGET-3} \citep{Springel05}. The simulations have a force softening length of 25 pc/$h$ and a high resolution dark matter particle mass of $2\times10^{4}$ ${\rm M}_{\odot}/h$. This corresponds to a resolved subhalo maximum circular velocity $V_{\rm max}=4.5\,\kms$. All of the simulations are cosmological zoom-in simulations \citep{Katz93,Onorbe14} with initial conditions generated by MUSIC \citep{Hahn11}, assuming a box size of 50 ${\rm Mpc}/h$, and a cosmology from Planck (\citeyear{Planck16}): $h = 0.6751, \,\Omega_{\Lambda} = 0.6879, \,{\rm and}\,\Omega_{\rm m}$ = 0.3156.

The host haloes in these simulations were selected to have halo masses at $z=0$ of  $M_{\rm vir} = 0.8-2\times10^{12}$ $\rm M_{\odot}$, in line with current estimates of the Milky Way halo mass. These haloes are also isolated such that they are the largest halo within 3 Mpc. The suite consists of 12 haloes run both with and without the disk implementation for a total of 24 simulations.

While a basic description of how the disk is implemented in the code can be found in \citet{SGK2017},  several modifications have been made to the properties of the galaxy potential in order to match precisely the Milky Way and its expected evolution.  Specifically, we mimic an exponential disk galaxy potential following \citet{Smith15}, who show that three summed Miyamoto-Nagai disks \citep{Miyamoto75} provide a good approximation to an exponential disk. We use this method to model the gas and stellar disks of the Milky Way.  We further include the bulge of the Milky Way as a Hernquist potential \citep{Hernquist90}.

The $z = 0$ values for the properties of the Milky Way bulge, stellar, and gas disks (including both masses and scale heights) were taken from \citet{JBH16} and \citet{McMillan17}. The disk is initialized at $z=3$ and the time evolution of the stellar mass is determined by tying the stellar mass growth to the halo mass growth using abundance matching from \citet{Behroozi13}. The evolution of the scale radii is then matched to CANDELS data from \citet{VDW14}. The gas mass is tied to the evolution of the cold gas fraction seen from CANDELS \citep{Popping15}, and the gas scale radius is fixed to be a constant multiple of the (time-evolving) stellar disk scale radius.  Finally, the bulge is modeled with a mass and scale radius fixed to that of the stellar disk so that the ratios are a constant as a function of time. A full description of the suite of simulations along with basic properties of their satellites will be given in Kelley et al. (in preparation).

\section{Subhalo counts and distributions}
\label{s:results}
Figure \ref{fig:subhalo_props} presents $V_{\rm peak}$ functions for subhaloes within 300 kpc of each host; $V_{\rm peak}$ is defined as the maximum of $ V_{\rm max}$ over all time, which is usually reached prior to subhalo infall at a distance of 1.5 to 7 virial radii from the host \citep{Behroozi14}. The shaded bands correspond to the full width of the distributions over all simulations.  Compared to the dark matter only (dmo) simulations (black), the disk simulations (magenta) show a factor of $\sim 50\%$ less substructure within this volume \citep[consistent with][]{SGK2017}.  Both sets of simulations begin to flatten at $\Vpeak \simeq 6 \,\kms$, which we take as our completeness limit.  As mentioned above, we are complete to current maximum circular velocities of $ V_{\rm max} \simeq 4.5\,\kms$.  

The horizontal lines in Figure~\ref{fig:subhalo_props} show the number of classical Milky Way satellites and the current count of all satellite galaxies known.  The band shows a range of estimates\footnote{All of these completeness corrections assume that the radial distribution of satellites follows the results of dmo simulations.  The corrections will {\em increase} if the relative depletion of central subhaloes from the galaxy potential is included.  In this sense, the corrections shown here are conservatively low.} for the total number of satellite galaxies after accounting for incompleteness and sky coverage limits \citep{Tollerud08,Kim17,Newton18}. We see that, based on counts, the classical satellites are consistent with sitting in haloes with $V_{\rm peak}$ $\geq 30 \,\kms$. The range of completeness-corrected satellites corresponds to $V_{\rm peak}$ between 8 and 18 $\kms$. The lower end of estimates ($\sim 100$) is more in line with the standard expectation for reionization quenching at $ V_{\rm peak} \simeq 20 \,\kms$.  The upper end of the range ($\sim 600$) would suggest the need to populate quite small haloes $V_{\rm peak} \simeq 8 \,\kms$, well below the atomic cooling limit.

As shown by \citet{SGK2017}, disk destruction is particularly important at small radii. 
In the right panel of Figure \ref{fig:subhalo_props}, we show the radial distribution of satellites with $V_{\rm peak} >6 \,\kms$ out to 100 kpc for both the dmo and dmo+disk simulations.  As before, the bands show the full width over all simulations and the solid lines show the medians.  The disk simulations retain very little substructure within 20 kpc.   

The vast majority of subhaloes have $z_{\rm peak}$ $\leq$ 3 (97\% in the disk simulations and 93\% in the dmo simulations). The average $z_{\rm peak}$ for surviving subhaloes  within 50 kpc is $\langle z_{\rm peak}\rangle$ = 0.77 for the disk runs  and $\langle z_{\rm peak} \rangle$ = 0.94 for the dmo runs.  This difference is due to the enhanced subhalo destruction caused by the disk, which preferentially destroys subhaloes that fall in early (see Kelley et al. in preparation).

\begin{figure}
	\includegraphics[width=0.48\textwidth, trim = 25 40 0 0]{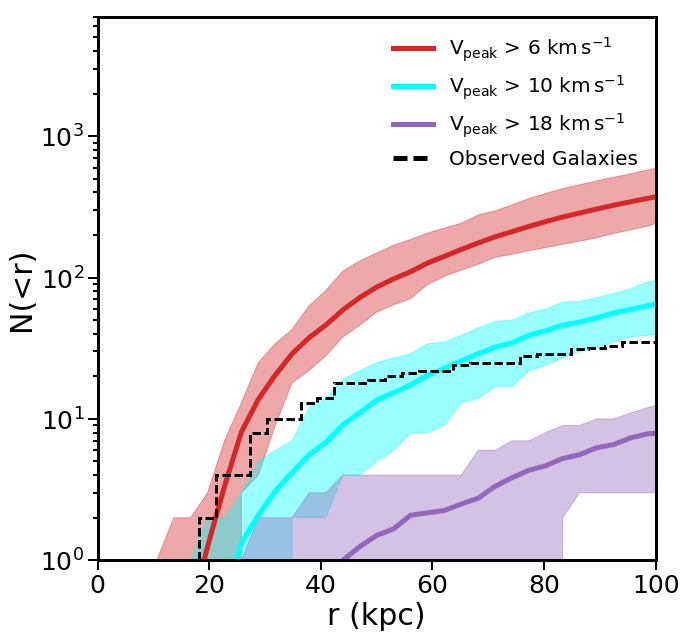}
	\caption[large_small_radii]{Radial distributions of subhaloes with $V_{\rm peak}>6, \,10$, and $18 \,\kms$ in the disk runs. The dashed line shows the radial distribution of known satellite galaxies, which is lower limit on the total.  Half of the sky has not been surveyed for ultrafaint dwarfs and the other half is incomplete at radii beyond $\sim 30$ kpc.  Still, the current satellite census is above the median subhalo counts at small radius unless we associate galaxies with the smallest subhaloes we resolve $ V_{\rm peak}> 6 \,\kms$.  The nominal reionization threshold would lead to an expectation close to the $18 \,\kms$ line (purple), which is far below the data.}
	\label{fig:large_small_radii} 
\end{figure}

\begin{figure*}
	\includegraphics[width=0.49\textwidth, trim = 0 0 0 0]{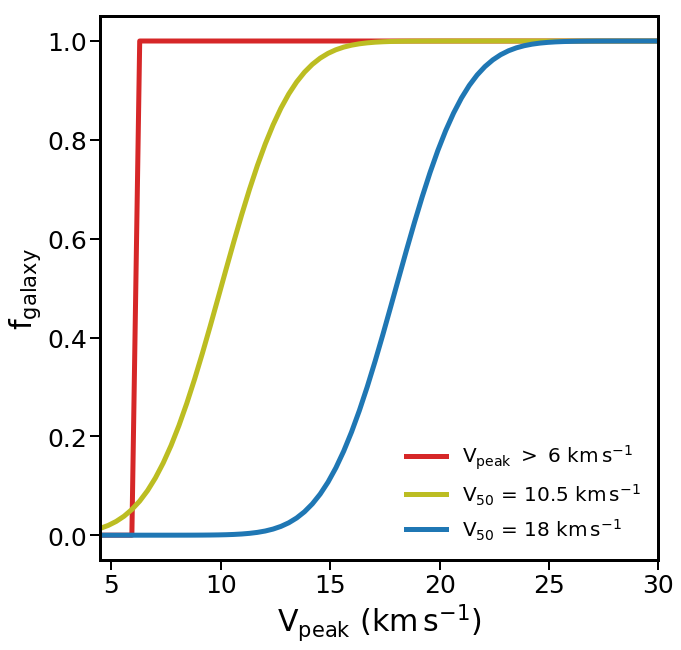}
	\includegraphics[width=0.49\textwidth, trim = 0 0 0 0]{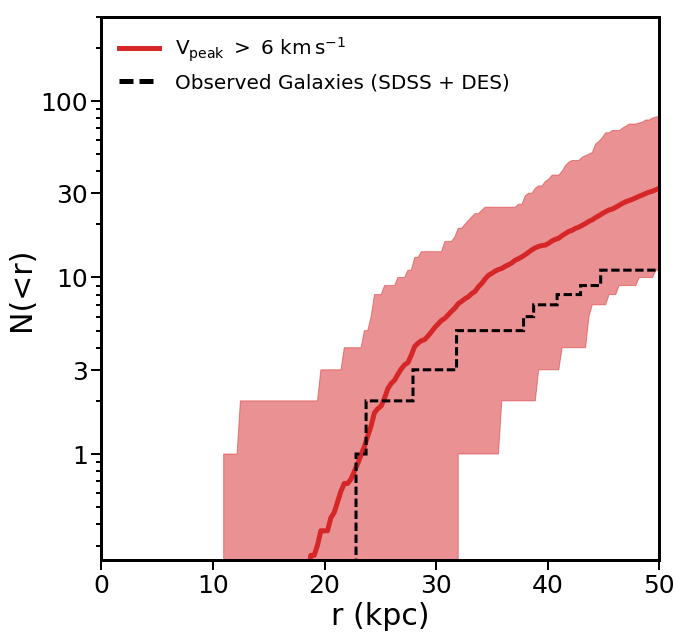}
    \includegraphics[width=0.49\textwidth, trim = 0 0 0 0]{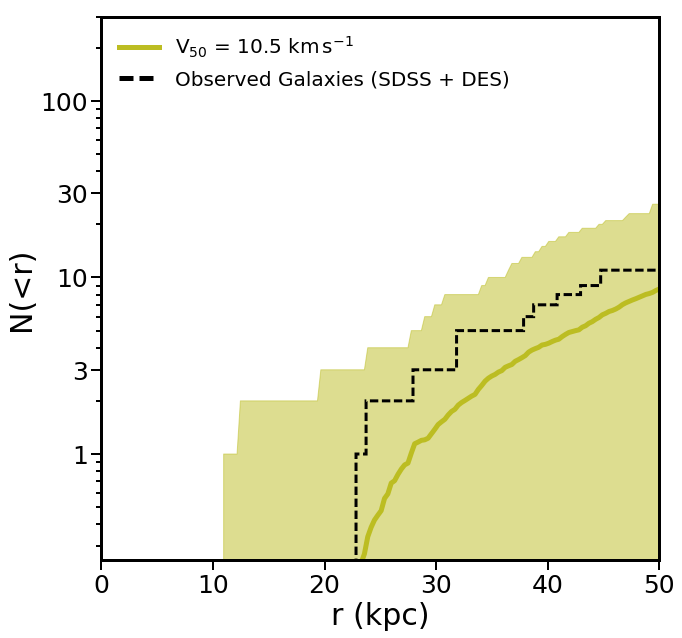}
    \includegraphics[width=0.49\textwidth, trim = 0 0 0 0]{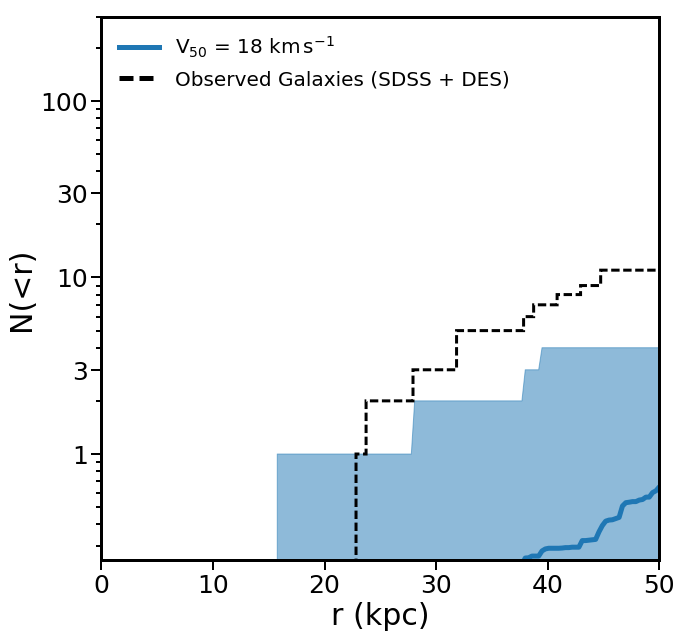}
	\centering
	\caption[radial_distribution]{\textit{Top-left}: The models used to calculate which haloes could potentially host a galaxy. The blue is a `conventional' model based on estimates of the dark fraction from hydrodynamic simulations. The dashed black lines in the other panels show the radial distribution of Milky Way satellites restricted to those within the SDSS and DES regions. These distributions are almost certainly incomplete at large radius, and thus represent lower limits. \textit{Top-right}: The predicted radial distribution of satellites within the SDSS plus DES regions assuming every halo with $\Vpeak \geq 6\,\kms$ forms a galaxy. \textit{Bottom-left}: The distribution after applying the $V_{50} = 10.5  \,\kms$ model. \textit{Bottom-right}: The distribution for the $V_{50} = 18 \,\kms$ model.   The predicted distributions come from mock observations within angular regions that mimic SDSS and DES coverage. The bands thus represent the full halo-to-halo scatter and scatter from anisotropy in the distribution of subhaloes.} 
	\label{fig:radial_distribution}
\end{figure*}

Figure \ref{fig:large_small_radii} shows only the (more realistic) disk simulations, now restricted to  the inner 50 kpc.  The three bands show radial distribution of subhaloes with $V_{\rm peak}>$ 6, 10, and $18 \,\kms$. Note that there are typically no subhaloes with $\Vpeak >18\,\kms$ that survive within 40 kpc. This is surprising given that we certainly know of satellite galaxies within 40 kpc of the Milky Way, and $\Vpeak \simeq 18\,\kms$ is close the conventional scale for reionization suppression where haloes begin to go dark.

The black dashed line shows the radial distribution of known satellite galaxies out to 100 kpc.  We know the census of satellites is incomplete both radially (due to luminosity incompleteness) and in area on the sky (less than $\sim 1/2$ of the sky has been covered in searches capable of finding ultrafaint galaxies). Nevertheless, the total count of satellite galaxies exceeds the median subhalo count at small radius for all but the $V_{\rm peak} > 6\,\kms$ sample (red). The $V_{\rm peak}>18\,\kms$ distribution, which is closest to the canonical reionization suppression scale, drastically under-predicts the number of known galaxies \citep[see][ for a list of galaxies within $R_{\rm vir}$]{Newton18}.

\begin{figure*}
	\includegraphics[width=0.49\textwidth, trim = 0 0 0 0]{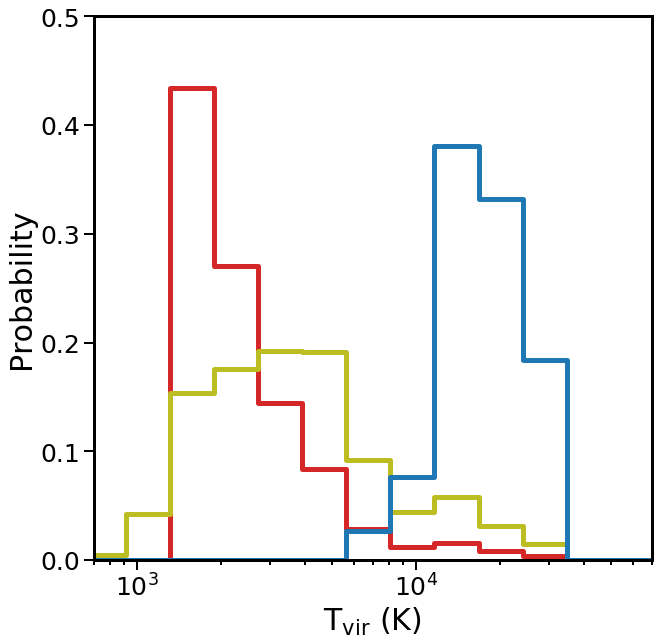}
	\includegraphics[width=0.5\textwidth, trim = 0 0 0 0]{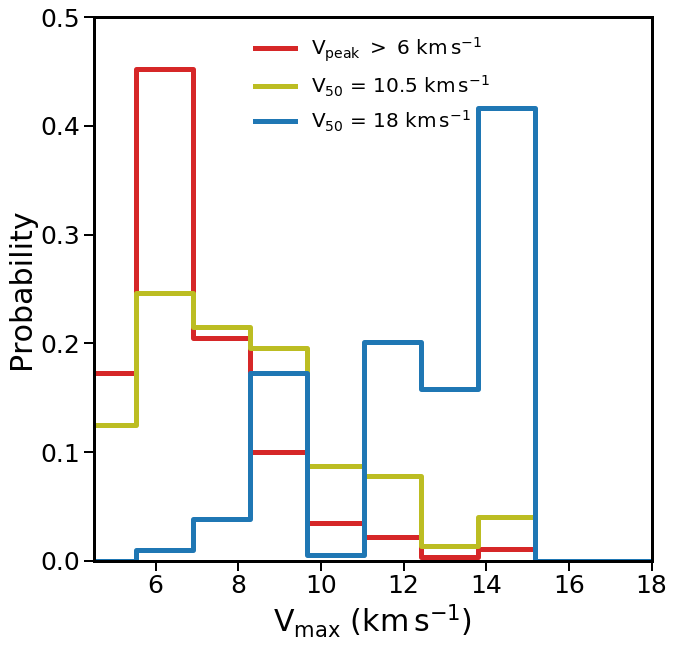}
	\centering
	\caption[V_peak_sims]{ \textit{Left:} The distribution of $T_{\rm vir}$ values for for the suhaloes populated by galaxies in the models shown in Figure \ref{fig:radial_distribution}.  Recall that the $V_{50} = 18\,\kms$ model (blue) drastically under-predicts Milky Way satellite counts, while the $V_{50} = 10.5\,\kms$ (yellow) and $\Vpeak > 6\,\kms$ (red) models produce adequate counts.  The consistent models all require galaxies to exist in haloes with virial temperatures of $\lesssim 4,000$K, well below the atomic cooling limit.   \textit{Right:} The $V_{\rm max}$ distributions at $z=0$ for subhaloes within 50 kpc of the Milky Way for the same models. The successful models peak below $7\,\kms$ today. }
\label{fig:V_peak_sims}
\end{figure*}

Figure \ref{fig:large_small_radii}  demonstrates that in order to account for the satellite galaxies of the Milky Way that are known to exist within $\sim 30$ kpc, we need to resort to populating haloes with $V_{\rm peak}$ values that are substantially lower than canonical values for reionization quenching that have been discussed in the literature.  These counts are known to be significantly incomplete at larger radius even in areas of the sky that have been covered by surveys like SDSS and DES.  Below, we present a somewhat more detailed exploration of what the ultrafaint Milky Way satellite population tells us about the low-mass threshold of galaxy formation.  

\subsection{Satellite Occupation Fractions}

Simulations that have explored how haloes go dark at low masses typically find that the fraction of haloes hosting a galaxy of any mass ($f_{\rm galaxy}$) drops towards zero smoothly below a characteristic value of $\Vpeak$ \citep[e.g.,][]{Sawala16b,Fitts18}.    
In order to allow for this expectation, we have explored a toy model where the fraction of haloes that host a galaxy of any mass varies smoothly from zero at small $\Vpeak$ to unity as $\Vpeak$ increases.  We specifically adopt a cumulative Gaussian, which allows for a characteristic scale ($V_{50}$) where $50\%$ of haloes become dark and a width ($\sigma$) that sets the sharpness  of the the transition from dark haloes to galaxy-hosting haloes:
\begin{equation}
f_{\rm galaxy}(>\Vpeak) = \frac{1}{2} \left[ 1 + \erf \left(\frac{\Vpeak - V_{50}}{\sqrt{2} \sigma }\right) \right].
\label{eq:f}
\end{equation}

The upper left panel of Figure~\ref{fig:radial_distribution} shows two models of this kind along with a simple threshold model (red) for comparison.  The blue line shows a conventional   model with $V_{50} = 18 \,\kms$, and $\sigma = 2.5$.  These  values are chosen to match the FIRE-2 results presented in \citet{Fitts18}, but they are typical of other results in the literature: haloes begin to go dark at $\Vpeak \simeq 25\,\kms$ and go completely dark by $\Vpeak \simeq 10\,\kms$.  A case that shifts the quenching scale a factor of $\sim 3$ lower in virial temperature is shown in yellow:  $V_{50} = 10.5 \,\kms$  with $\sigma = 2.5$. 

In order to compare the predictions of these simple models to the observed population of Milky Way satellites, we take into account sky coverage completeness and account for the fact that subhalo populations are anisotropic.  We restrict ourselves to only satellites within either the SDSS or DES footprints, both of which have well-defined completeness areas.  We make no allowance for luminosity/volume incompleteness in order to be conservative. Using Equation \ref{eq:f} we assign a galaxy to each subhalo probabilistically for all 12 of our disk simulations; and repeat this procedure 100 times counting `galaxies' as a function of radius within mock survey areas. The SDSS and DES survey regions are approximated as three cones with the areas of the two contiguous SDSS fields and the DES field, and their orientations are fixed relative to one another to match the surveys.  Each iteration uses a different orientation of the cones.  Note that unlike the real DES and SDSS fields, we orient the survey cones randomly with respect to the disk planes. We do this because it increases our statistics and because we find that the disk does not introduce any significant asymmetry (in fact, it sphericalizes the subhalo distributions compared to the dmo runs). 

 The top-right panel in Figure \ref{fig:radial_distribution} shows the results of this exercise for our simple $V_{\rm peak}$ $\geq 6 \,\kms$ threshold model. The shaded band includes the full scatter over all simulations and survey orientations, with the solid line showing the average. The bottom-left and bottom-right panels show the full distributions (minimum and maximum) of the $V_{50} = 10.5\,\kms$ and $V_{50} = 18 \,\kms$ models respectively. For comparison, the galaxies in SDSS and DES are shown as a black histogram.
 Table 1 lists all satellite galaxies within 50 kpc of the Milky Way.  We include only the satellites that sit within the SDSS or DES footprints\footnote{Currently we know of about 58 satellites around the Milky Way; however, this includes many candidate ultrafaints from surveys such as DES, MagLiteS, and PanSTARRS that have yet to be spectroscopically confirmed.} in Figure \ref{fig:radial_distribution}.  The current census is incomplete at large radius so the dashed lines in Figure \ref{fig:radial_distribution} are lower limits.

As is clear from Figure~\ref{fig:radial_distribution}, the $\Vpeak$ $\geq 6\,\kms$ model matches the observed distribution the best at small radius (where incompleteness matters least).  This is surprising since these haloes are far lower-mass than those naively expected to host galaxies.  In contrast, the most well-motivated model, with $\rm V_{50} = 18 \,\kms$, fails to form enough galaxies in the inner regions to match the number of galaxies already observed.  In fact, the average is less than one galaxy out to 50 kpc.
 The $V_{50} = 10.5 \,\kms$ model produces a distribution that is consistent with the observed counts, primarily because it has a tail of non-zero occupation that stretches down to low-mass haloes.

Another way to see how surprisingly low-mass the required haloes are is to look at the distribution of virial temperatures for the haloes in these models.  For this, we use:
\begin{equation}\label{eqn:Tvir}
T_{\rm vir} \simeq 10^4 \rm{K} \left(\frac{\Vpeak}{16.3 \,\kms} \right)^2 , \\
\end{equation}
which follows from $T_{\rm vir} = \mu m_{\rm p} c_{\rm g}^2 / k_b$ with $c_{\rm g} =  V_{\rm max}/\sqrt{2}$ and $\mu = 0.62$, as implied by a 30\% mass fraction in helium.  The left panel of Figure \ref{fig:V_peak_sims} shows the $T_{\rm vir}$ distribution for the galaxy-populated subhaloes within $50$ kpc for each of the three models discussed in relation to Figure \ref{fig:radial_distribution}.  The two models that produce consistent Milky Way satellite populations all require that we populate subhaloes with $T_{\rm vir} < 4,000$ K.  This is well below the atomic cooling limit and the canonical reionization quenching scale.

What do the dark matter haloes of these galaxies look like today?  The right panel of Figure \ref{fig:V_peak_sims} shows the distribution of $V_{\rm max}$ values (at $z=0$) of the populated subhaloes in each of our three example models.  Both of the consistent models (yellow and red) populate haloes with $V_{\rm max} \simeq 5 - 10\,\kms$ today, with a small tail of maximum circular velocities out to $\sim 15\,\kms$.  While we cannot measure the maximum circular velocities of Milky Way dwarf satellites directly, we can measure the circular velocity at the half-light radius, $V_{c}(r_{1/2})$. This represents a lower limit on the true $V_{\rm max}$ of the halo. Values of $ V_{c}(r_{1/2})$ for the Milky Way dwarfs within 50 kpc are listed in Table 1. Interestingly, of the 18 galaxies within 50 kpc (if surveys like MagLiteS and PanSTARRS are included) 3 have $V_{c}(r_{1/2}) > 10 \,\kms$, which is already larger than typical $V_{\rm max}$ values in the successful models.  The rest have circular velocities at $r_{1/2}$ that are at least consistent with those expected.  Determining $V_{\rm max}$ estimates based on these measurements will requite density profile priors to extrapolate from $r_{1/2}$ to the peak of the rotation curve at $r_{\rm max}$ \citep[e.g.][]{MBK2011}.  Such an exploration is beyond the scope of this paper, but for context, a $V_{\rm max} \simeq 15 ~\kms$ ($7 ~\kms$) halo typically has $r_{\rm max} \simeq 1.5$ kpc (500 pc) \citep[e.g.][]{SGK2014}. This is a significant extrapolation for most ultrafaint dwarfs.

\begin{table*}
  \caption{Table of galaxies within 50 kpc of the Milky Way.  Galaxies within the SDSS and DES footprints (with well-understood completeness areas) are listed above the solid line. Galaxies below the horizontal line are other galaxies known to exist within 50 kpc. We provide Galacto-centric radius, stellar velocity dispersion, implied circular velocity at the half-light radius, and the half-light radius.  Note that $ V_{c}(r_{1/2})$ is a lower limit on the current value of $V_{\rm max}$.}
\centering 
\begin{tabularx}{\textwidth}{lXXXXXXX}
\hline
\hline  
Name &$\rm D_{GC}$ & $\rm \sigma_{star}$ & $V_{c}(r_{1/2}) $ & $r_{1/2} $ & source & Ref.$^a$ & $\rm \sigma_{star}$ Ref.$^b$\\ 
  &[$\rm kpc$] & $[\kms]$ & $[\kms]$ & $[{\rm pc}]$ & & --& \\
\hline 
Galaxy & (1) & (2) & (3) & (4) & & & \\
\hline 
DES J0225+0304 & 22 & -- & -- & 18.5 & DES & (1) & -- \\
Tucana III & 23 & $\leq$ 1.2 & $\leq$ 2.1 & 44 & DES & (2),(3) & (15)\\
Segue I & 28 & $3.7^{+1.4}_{-1.4}$ & 6.4 & 29 & SDSS DR6 & (4) & (16)\\
Cetus II & 32 & -- & -- & 17 & DES & (2),(3) & -- \\
Reticulum II & 32 & 3.3 $\pm$ 0.7 & 5.7 & 32 & DES & (2),(3) & (17)\\ 
Ursa Major II & 38 & 6.7 $\pm$ 1.4 & 11.6 & 149 & SDSS DR4 & (5) & (18)\\
Bootes II & 39 & 10.5 $\pm$ 7.4& 18.2 & 51 & SDSS DR5 & (6) & (19)\\
Segue II & 41 & $\leq$2.2 & 3.8 & 35 & SDSS DR7 & (7) & (20)\\
Willman I & 43 & 4.0 $\pm$ 0.8 & 6.9 & 25 & SDSS DR2 & (8) & (21)\\
Coma Berenices & 45 & 4.6 $\pm$ 0.8 & 8.0 & 77 & SDSS DR5 & (4) & (18)\\
Tucana IV& 45 & -- & -- & 127 & DES & (2),(3) & -- \\
\hline  

Sagittarius I & 18 & 9.6 $\pm$ 0.4 & 16.62 & 2587 & Classical & (9) & (22) \\
Draco II& 20 & $\leq$ 8.4 & $\leq$ 14.5 & 19 & Pan-STARRS & (10) & (23)\\
Hydrus 1 & 20 & $2.69^{+0.51}_{-0.43}$ & 4.7 & 53 & DECam & (11) & -- \\
Carina III& 29 & -- & -- & 30 & MagLiteS & (12) & -- \\
Triangulum II& 30 & $\leq$ 3.4 & $\leq$ 5.9 & 34 & Pan-STARRS & (13) & (24)\\
Carina II& 37 & -- & -- & 91 & MagLiteS & (12) & -- \\
Pictoris II& 45 & -- & -- & 46 & MagLiteS & (14) & -- \\

\hline
\label{table:galaxy_props}
\end{tabularx}
\begin{tablenotes}
	\item[a] $^{a}$ (1) \citet{Luque17} (2) \citet{ADW15}, (3) \citet{Koposov15}, (4) \citet{Belokurov07}, (5) \citet{Zucker06}, (6) \citet{Walsh07}, (7) \citet{Belokurov09}, (8) \citet{Willman05}, (9) \citet{Ibata14}, (10) \citet{Laevens15a}, (11) \citet{Koposov18}, (12) \citet{Torrealba18},  (13) \citet{Laevens15b}, (14) \citet{ADW16}
    \item[b] $^{b}$ (15) \citet{Simon17}, (16) \citet{Simon11}, (17) \citet{Simon15}, (18) \citet{Simon07}, (19) \citet{Koch09}, (20) \citet{Kirby13}, (21) \citet{Willman11}, (22) \citet{Bellazzini08}, (23) \citet{Martin16}, (24) \citet{Kirby17}
\end{tablenotes}
\end{table*}

\section{Caveats and Solutions}
\label{s:caveats}

\subsection{LMC Bias}

One of the key components in understanding the present-day dwarf population of the Milky Way has been contributions from DES. DES has revealed a large number of ultrafaint galaxies in the vicinity of the sky near the LMC, which immediately leads to the question of whether most, if not all, of these ultrafaints fell in with the LMC. This scenario is not difficult to imagine, as LMC-mass haloes could easily host tens of ultrafaint dwarf satellites themselves \citep{Deason15,Dooley17}. Several works have investigated which of the DES dwarfs can reasonably be associated with the LMC. For example, \citet{Sales17} suggest that of the DES dwarfs within 50 kpc, only Tucana IV can be potentially associated with the LMC. However, \citet{Jethwa16} suggest that three of the DES dwarfs within 50 kpc can be reasonably associated with the LMC, with Tucana III having a $\geq$ 50\% probability of being associated with the LMC, and Reticulum II and Tucana IV having a $\geq$ 70\% probability of association.

Five of the 18 dwarf galaxies within 50 kpc of the Milky Way were discovered in DES fields.  If we assume that all of these are associated with the LMC and therefore are not fairly compared to our simulations (none of which have an LMC-like system at a similar distance), we can compare our expectations to a total of 13 non-LMC dwarfs.  From Figure \ref{fig:large_small_radii}, we see that this would allow us to more easily populate only $\Vpeak > 10\,\kms$ dwarfs with galaxies.  These are still below the atomic cooling limit, but not as drastically as those in the preferred models discussed above.  If the LMC is indeed this critical to a full understanding of the satellite population of the Milky Way, this justifies a dedicated program to simulate Milky Way haloes with targeted LMC-like subhaloes in the future.

\subsection{Numerical Disruption}

Another potential explanation for the low mass scale required to explain the radial distribution of satellites is that subhalo disruption at small radii is dominated by numerical error. If a factor of $\sim 5$ more $\Vpeak \simeq 20 \,\kms$ subhaloes survive at small radii than we are capturing, then the primary concern would go away.  While our simulations pass basic convergence tests in the presence of disk potentials down to masses $\sim 1000$ times smaller than this $\Vpeak$ value \citep[see][who have slightly worse resolution than our simulations]{SGK2017}, this is not a guarantee that we (or other simulations) are free from numerical error \citep{vdb18b}.

There is a fairly extensive body of literature suggesting that cuspy dark matter haloes are never completely disrupted and that a tiny cusp always survives even if systems lose $>$ 99 \% of their initial mass \citep{Goerdt07,Penarrubia10,vdb18a}. Of particular relevance in determining if our simulations could be subject to substantial numerical disruption is the work of \citet{vdb18b}. They suggest that many haloes are artificially disrupted at far higher halo masses than would be naively assumed by convergence studies of halo mass functions (which is how we set our resolution in this work).  They also provide criteria for determining if a subhalo is safe from the effects of numerical disruption. For a circular orbit at 10\% of the virial radius, a subhalo needs to be simulated with about $10^6$ particles and a softening length of 0.003 times the scale radius of the subhalo.  These criteria vary strongly with the tidal field the subhalo experiences. 

The criteria suggested in \citet{vdb18b} would place the mass resolution for our simulations at $\Vpeak \simeq 30 \,\kms$. We identify a severe discrepancy with the conventional reionization threshold at a $\Vpeak$ value only slightly smaller this, but the tidal field should be even stronger in the disk simulations, so the resolution criteria may be even more restrictive. 

\begin{figure}
	\includegraphics[width=0.99\linewidth, trim = 0 0 0 0]{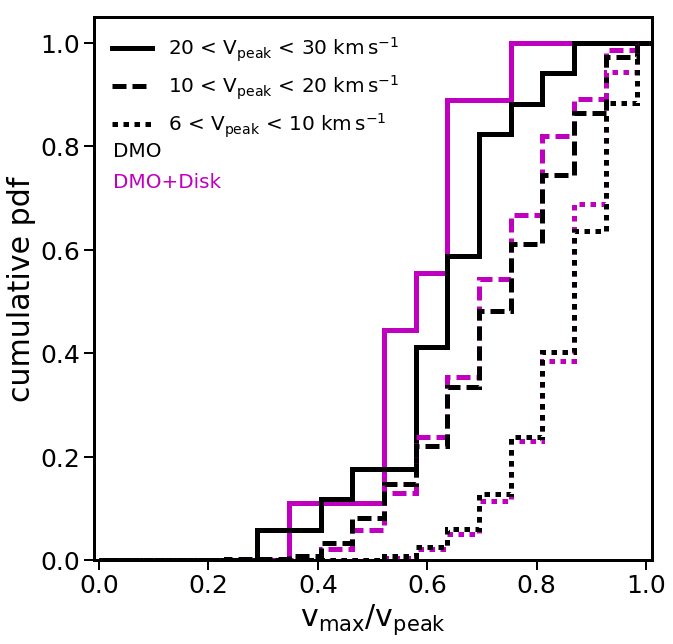}
	\caption[subhalo_stripping]{The effect of stripping on surviving subhaloes from both the dmo (black) and disk (magenta) simulations, represented as the ratio between present day $V_{\rm max}$ and $\Vpeak$. The different line styles represent different bins in $\Vpeak$. Interestingly, the stripping is only different in the highest-mass bin, for subhaloes between 20 and 30 $\,\kms$. However, the stripping is similar for the dmo and disk runs in other mass bins. Note that our resolution limit does not allow us to capture subhaloes much smaller than $V_{\rm max}=4.5\,\kms$, which explains why the higher-mass subhalo bins can be tracked to smaller $V_{\rm max}/\Vpeak$ ratios.}
	\label{fig:subhalo_stripping}
\end{figure}

A related question is by how much our surviving subhaloes are stripped after infall. We explore this in Figure \ref{fig:subhalo_stripping}, where we plot the differences in stripping for the surviving haloes in both the dmo and dmo+disk simulations within 50 kpc. Interestingly, the only difference between the two sets of simulations is seen in the highest mass bin (20 $\kms$ to 30 $\kms$), where subhaloes in the disk simulation are stripped on average 5\% more than in the dmo simulations. For subhaloes with $ V_{\rm peak}$ $\leq 20 \,\kms$, the difference in stripping between the dmo and disk simulations is much smaller, just about 1\% on average. This implies that the disk is causing significant damage to these subhaloes, driving their masses below our resolution limit, rather than simply lowering their masses and allowing them to survive.  After their masses drop below our resolution limit, they are `destroyed' in our catalogs only because we cannot track them anymore. This means that haloes in the $\Vpeak = 10$ to 20 $\kms$ range will drop out of our catalogs at $V_{\rm max}/\Vpeak \approx 0.3$, which is roughly where the relevant histogram in Figure \ref{fig:subhalo_stripping} drops to zero, and will be subjected to 90\% to 99\% mass loss from stripping.  There is no indication that the high-mass group of haloes (solid lines) has a tail that is boosted with respect to the lower-mass bins (dashed), as might be expected if there was a catastrophic numerical threshold at $\Vpeak \simeq 20 ~\kms$, potentially indicating that our simulations are not subject to numerical disruption above our assumed completeness. However, we stress that more work is needed to understand the results of \cite{vdb18b} in the context of cosmological simulations of the kind employed here.

\subsection{Star formation in low-mass haloes}

Another option is that galaxies are indeed forming in haloes that are lower mass than conventional models have suggested. If this is the case, star formation in local ultrafaint dwarfs could have proceeded by molecular cooling, as expected for Pop III formation in mini-haloes \citep{Bromm04,JBH15}. If the temperature distributions presented Figure \ref{fig:V_peak_sims} reflect a physical reality, ultrafaints must form in the kind of haloes usually associated with the first stars.

Another possibility is that reionization happened later than is often assumed in the volume surrounding the Milky Way.  
Naively, we expect star formation to proceed in galaxies at high redshift if they have circular velocities above the atomic cooling scale prior to reionization  $V_{\rm max} > 16 ~\kms$.  After reionization, star formation is assumed to shut down unless the halo is above the suppression scale \citep[typically 20 to 30 $\kms$; see][for a recent discussion]{Fitts17}.  If reionization happened late, we would have more small haloes in place earlier, perhaps increasing the global likelihood for galaxy formation at a fixed halo mass.

Recently, Munshi et al. (in preparation) have used ChaNGa simulations to show that the relationship between haloes and galaxies at very low masses is sensitive to the star-formation threshold density adopted.  Specifically, lower thresholds (e.g. $100$ cm$^{-3}$ rather than $1000$ cm$^{-3}$) give more star formation in smaller haloes.  While several state-of-the art simulations have shown that galaxy properties are insensitive to the adopted star formation threshold \citep[see][and references therein]{Hopkins17}, this is only true in larger galaxies where star formation self-regulates via feedback.  In the smallest haloes ($\Vpeak \lesssim 20 \,\kms$), star formation is regulated by the external ionizing field and thus can be more sensitive to adopted threshold (Munshi et al., in preparation).  Future work in this direction may provide important physical insights into how we might naturally form faint galaxies in $\Vpeak \simeq 6-10 ~\kms$ haloes.

\section{Conclusion} \label{s:conclusion}

In this work, we utilized a new suite of 12 cosmological simulations of Milky Way-like haloes that include a central disk potential (Kelley et al., in preparation).   The haloes were chosen to match the mass of the Milky Way ($ M_{\rm vir}=0.8-2\times10^{12}$ $\rm M_{\odot}$). The galaxy is modeled as an evolving disk + bulge potential that grows to match the Milky Way at $z=0$.  The inclusion of the galaxy drastically affects the subhalo abundance at small radii, as shown in Figure \ref{fig:subhalo_props}, which is consistent with past work based on hydrodynamic simulations \citep{Brooks14,Zhu2016,Wetzel2016,SGK2017,Graus18}. 

We compared the subhalo distributions at small radius to the current census of galaxies that exist within 50 kpc of the Milky Way disk, most of which are ultrafaint galaxies discovered in digital sky surveys \citep[e.g.][]{Willman05,Zucker06,ADW15,Koposov18}.  Even though our census of small galaxies is likely incomplete, we require very low-mass haloes to host galaxies in order account for all of the presently-known galaxies (see Figure~\ref{fig:large_small_radii}).  In particular, we need haloes as small as $ V_{\rm peak} = 6 \,\kms$ to host  ultrafaint galaxies.  These systems have virial temperatures as low as $\sim 1,500$ K, which is not only well below the typical scale where reionization is expected to suppress galaxy formation ($\Vpeak \simeq 20 ~\kms$ and $T_{\rm vir} \simeq 15,000$ K) but also much smaller than the atomic cooling limit ($\Vpeak \simeq 16 ~\kms$ and $T_{\rm vir} = 10,000$ K). 

We explored these results by mock-observing our Milky Way haloes over regions that mimic DES and SDSS fields using toy models that allow the fraction of haloes that host galaxies to vary from 0 to 1 at a characteristic $\Vpeak$ scale.  As shown in Figure \ref{fig:V_peak_sims}, the models that work populate haloes with $\Vpeak$ values between $6$ and $16 ~\kms$ ($T_{\rm vir} =  1,500 - 10,000$ K), much smaller than would be conventionally expected.  It is important to note that any new discoveries of ultrafaint dwarf galaxies close to the Milky Way would only increase the need to populate very small haloes with galaxies. 

There are at least three possibilities that could change these conclusions.  First, several of the ultra-faint `galaxies' we include in our analysis (Table 1) could be misclassified star clusters.  Ultra-faint galaxies are differentiated from star clusters by inhabiting dark matter haloes  \citep{WS12}. If some fraction of the satellites in our comparison do not have dark haloes then the implied threshold $\Vpeak$ for galaxy formation would increase accordingly.  

A second possibility is that many, if not most dwarfs that were discovered by DES were brought in with the LMC. If this is the case, it would bias the Milky Way dwarf population to be over-abundant compared to what is typical for haloes of the Milky Way's mass. Of the 18 total dwarf galaxies within 50 kpc of the Milky Way, five were discovered in DES.  Figure \ref{fig:large_small_radii} shows that if all five of these were removed from the comparison the need to populate very low-mass subhaloes would be lessened, but that there would still be tension unless we allow for most $\Vpeak \simeq 10 ~\kms$ haloes to host galaxies.  Even in this fairly conservative scenario (which ignores sky coverage incompleteness) we require galaxy formation below the atomic cooling limit. 

A third possibility is that much of the destruction we see is a numerical artifact \citep{vdb18a}. By conventional convergence-test standards we appear to be well-resolved down to $\Vpeak = 6 ~\kms$ and certainly to $\Vpeak = 10 ~\kms$; however, by the criteria described in \citet{vdb18b}, we could be affected by numerical issues at the critical scale $\Vpeak = 20 ~\kms$.  If so, then ultrafaint galaxies may reside within halos with $V_{\rm peak}$ above the canonical reionization suppression scale, and the numerical challenge facing Milky Way satellite modelers will become quite significant.

Aside from the caveats discussed above, our results suggest that haloes well below the atomic cooling limit host ultrafaint galaxies.   One implication of such a scenario is that we would expect $\sim 1000$ such systems within 300 kpc of the Milky Way (see the left panel of Figure ~\ref{fig:subhalo_props} at $\Vpeak \simeq 7 ~\kms$ and the last row of Table 1 in \citealt{Kim17}).  This count will be testable with LSST and is significantly higher than previous completeness-correction estimates, which had relied on dark-matter only simulations.  Whatever the answer, the results presented here motivate renewed efforts to understand galaxy formation in the the smallest haloes and the effect of the near-field environment on their evolution. 

\section{Acknowledgments} \label{s:Acknowledgements}

The authors would like to thank Josh Simon for providing the dispersion velocities of dwarf satellites used in this work, and Alyson Brooks, Alex Drlica-Wagner, and Denis Erkal for useful discussions.  ASG, TK, CQ, and JSB were  supported by NSF AST-1518291, HST-AR-14282, and HST-AR-13888.  MBK acknowledges support from NSF grant AST-1517226 and CAREER grant AST-1752913 and from NASA grants NNX17AG29G and HST-AR-13888, HST-AR-13896, HST-AR-14282, HST-AR-14554, HST-AR-15006, HST-GO-12914, and HST-GO-14191 from the Space Telescope Science Institute, which is operated by AURA, Inc., under NASA contract NAS5-26555. Support for SGK was provided by NASA through the Einstein Postdoctoral Fellowship grant number PF5-160136 awarded by the Chandra X-ray Center, which is operated by the Smithsonian Astrophysical Observatory for NASA under contract NAS8-03060. This work used computational resources of the Texas Advanced Computing Center (TACC; http://www.tacc.utexas.edu), the NASA Advanced Supercomputing (NAS) Division and the NASA Center for Climate Simulation (NCCS), and the Extreme Science and Engineering Discovery Environment (XSEDE), which is supported by National Science Foundation grant number OCI-1053575.




\bibliographystyle{mnras}
\bibliography{radial_distribution_paper_refs.bbl} 








\bsp	
\label{lastpage}
\end{document}